\documentclass{moriond}

\def\be{\begin{equation}}
\def\ee{\end{equation}}
\def\bea{\begin{eqnarray}}
\def\eea{\end{eqnarray}}

\usepackage{amsmath,amssymb}
\usepackage{bm}

\newcommand{\Photo}{}
\newcommand{\ud}{\mathrm{d}}
\newcommand{\ui}{\mathrm{i}}
\newcommand{\ue}{\mathrm{e}}

\begin{document}
\title{
  Conservative Dynamics of Binary Systems of
  Compact Objects \\at the Fourth Post-Newtonian Order}

\author{ Laura BERNARD }

\address{CENTRA,
  Departamento de F\'{\i}sica, Instituto Superior T{\'e}cnico -- IST,
  \\Universidade de Lisboa -- UL, Avenida Rovisco Pais 1, 1049 Lisboa,
  Portugal}

\author{ Luc BLANCHET }

\address{$\mathcal{G}\mathbb{R}\varepsilon{\mathbb{C}}\mathcal{O}$,
  Institut d'Astrophysique de Paris,\\ UMR 7095, CNRS, Sorbonne
  Universit{\'e}s \& UPMC Univ Paris 6,\\ 98\textsuperscript{bis}
  boulevard Arago, 75014 Paris, France}

\author{ Alejandro BOH{\'E} }

\address{Max Planck Institute for Gravitational Physics (Albert
  Einstein Institute), \\Am Muehlenberg 1, 14476 Potsdam-Golm, Germany}

\author{ Guillaume FAYE }

\address{$\mathcal{G}\mathbb{R}\varepsilon{\mathbb{C}}\mathcal{O}$,
  Institut d'Astrophysique de Paris,\\ UMR 7095, CNRS, Sorbonne
  Universit{\'e}s \& UPMC Univ Paris 6,\\ 98\textsuperscript{bis}
  boulevard Arago, 75014 Paris, France}

\author{ Sylvain MARSAT }

\address{Max Planck Institute for Gravitational Physics (Albert
  Einstein Institute), \\Am Muehlenberg 1, 14476 Potsdam-Golm, Germany}

\maketitle\abstracts{ We review our recent derivation of a Fokker
  action describing the conservative dynamics of a compact binary
  system at the fourth post-Newtonian (4PN) approximation of general
  relativity. The two bodies are modeled by point particles, which
  induces ultraviolet (UV) divergences that are cured by means of
  dimensional regularization combined with a renormalization of the
  particle's wordlines. Associated with the propagation of wave tails
  at infinity is the appearance of a non-local-in-time conservative
  tail effect at the 4PN order in the Lagrangian. In turn this implies
  the appearance of infrared (IR) divergent integrals which are also
  regularized by means of dimensional regularization. We compute the
  Noetherian conserved energy and periastron advance for circular
  orbits at 4PN order, paying special attention to the treatment of
  the non-local terms. One ambiguity parameter remaining in the
  current formalism is determined by comparing those quantities,
  expressed as functions of the orbital frequency, with self-force
  results valid in the small mass ratio limit.}

\section{Introduction}

Inspiraling and merging black-hole binary systems are the most common
sources of gravitational waves detectable by ground or space-based
laser interferometric detectors.~\cite{BuonSathya15} Banks of
extremely accurate replica of theoretical templates are a compulsory
ingredient of a successful data analysis for these detectors --- both
on-line and off-line. In the early inspiral phase, the post-Newtonian
(PN) approximation of general relativity should be pushed to extremely
high order. Furthermore, high accuracy comparison and matching of PN
results are performed with numerical relativity computations
appropriate for the final merger and ringdown phases. In this context,
we have undertaken the derivation of the equations of motion for
binary systems of compact (non-spinning) objects at the 4PN
order. Solving this problem is of great importance for various
applications, most notably numerical/analytical self-force
comparisons~\cite{ALBSW15} and effective-one-body
calculations,~\cite{BuonD99} and paves the way to the determination of
physical observables in the radiation field such as the orbital phase
at the 4PN order beyond the Einstein quadrupole formalism.

After the introduction by Lorentz \& Droste~\cite{LD17} of the
perturbative PN scheme for solving the Einstein field equations (EFE)
for weakly gravitating, slowly moving sources, it was further explored
in several historical works, including the famous paper on the motion
of $N$ planets at the 1PN order by Einstein, Infeld \&
Hoffmann.~\cite{EIH} In the 1980s, the PN scheme was successfully
applied to the derivation of the equations of motion of compact
binaries up to the 2.5PN order, where radiation reaction effects first
appear,~\cite{Dhouches} which put an end to the radiation reaction
controversy raging at the time.~\cite{Ehletal76} The 3PN dynamics was
tackled in the 2000s with the help of various methods, and the 4PN
order has been investigated since the early 2010s.

After first partial results obtained by means of the effective field
theory (EFT)~\cite{FS4PN,FStail} and the Arnowitt-Deser-Misner (ADM)
formalism,~\cite{JaraS13} the important effect of gravitational wave
tails at the 4PN order was included into the ADM
Hamiltonian.~\cite{DJS14,DJS16} This allowed a better control of the
IR divergences and the completion of the full 4PN dynamics, in spite
of the appearance of one unfixed numerical constant which could only
be set by comparison with self-force calculations. We report here on
our alternative approach,~\cite{BBBFMa,BBBFMb,BBBFMc} based on the
construction of a Fokker Lagrangian in harmonic coordinates, and whose
end result is physically equivalent to the one of the ADM
Hamiltonian formalism.~\cite{JaraS13,DJS14,DJS16}

\section{Fokker action for post-Newtonian sources}

The two compact objects are represented by means of non-spinning,
structureless particles, with masses $m_A$ and trajectories
$y_A^\mu(t)=(c\, t,\bm{y}_A(t))$ (with $A=1,2$), where $c$ represents
the speed of light. The corresponding matter action reads
\begin{equation}\label{Sm}
S_{m} = - \sum_A m_A c^{2} \int \ud t \sqrt{-(g_{\mu\nu})_A
  \,v_A^{\mu}v_A^{\nu}/c^{2}}\,,
\end{equation}
with $v_A^\mu=\ud y_A^\mu/\ud t=(c,\bm{v}_A)$. The time-dependent tensor
$(g_{\mu\nu})_A$ stands for the metric evaluated at the location of the
particle $A$. We deal with the divergences arising there by means of
dimensional regularization. On the other hand, the gravitational sector is
described by the Einstein-Hilbert action in Landau-Lifshitz form with the
usual harmonic gauge-fixing term
\begin{equation}\label{Sg}
S_{g} = \frac{c^{3}}{16\pi G} \int \ud^{4}x \, \sqrt{-g} \left[
  g^{\mu\nu} \left( \Gamma^{\rho}_{\mu\lambda}
  \Gamma^{\lambda}_{\nu\rho} - \Gamma^{\rho}_{\mu\nu}
  \Gamma^{\lambda}_{\rho\lambda} \right) -\frac{1}{2}
  g_{\mu\nu} g^{\alpha\beta} g^{\rho\sigma}\Gamma^{\mu}_{\alpha\beta}\,
\Gamma^{\nu}_{\rho\sigma} \right] \,,
\end{equation}
where $G$ is the Newton constant, $g={\rm det}\,g_{\mu\nu}$, and
$\Gamma^{\rho}_{\mu\nu}$ stands for the Christoffel symbols.

The gravitational action can be written in terms of the deviation of
the gothic metric from the inverse flat metric
$\eta^{\mu\nu}=\text{diag}(-1,1,1,1)$, namely
$h^{\mu\nu}=\sqrt{-g}g^{\mu\nu}-\eta^{\mu\nu}$. The action appears
then as an infinite non-linear power series in $h$, in which indices
on $h$ and on partial derivatives $\partial$ are lowered and raised
with the Minkowski metric $\eta$. The Lagrangian density
$\mathcal{L}_g$ can take various forms, obtained from each other by
integrations by parts. For our purpose, we adopt the form that starts
at quadratic order by terms like $\sim h\Box h$, i.e., the
``propagator'' form, with
$\Box=\eta^{\rho\sigma}\partial^2_{\rho\sigma}$ denoting the flat
d'Alembertian operator. Therefore, the general structure of our
Lagrangian density is $\mathcal{L}_g\sim h\Box h+h \partial h\partial
h + h\,h\partial h\partial h + \cdots$.

The harmonic gauge fixed action yields the following ``relaxed'' EFE:
\begin{align}
\Box h^{\mu\nu} &= \frac{16\pi G}{c^4}
\tau^{\mu\nu}\,, & \tau^{\mu\nu} &\equiv \vert g\vert T^{\mu\nu} +
\frac{c^4}{16\pi G}\Sigma^{\mu\nu}[h, \partial h, \partial^2
  h]\,.\label{EFE}
\end{align}
The quantity $\tau^{\mu\nu}$ denotes the pseudo stress-energy tensor of the
matter and gravitational fields, with
$T^{\mu\nu}=\frac{2}{\sqrt{-g}}\delta S_m/\delta g_{\mu\nu}$. The
gravitational source term $\Sigma^{\mu\nu}$ is at least quadratic in $h$ or
its first and second derivatives. Those wave-like equations have the same
Green function as in harmonic gauge, although the harmonicity conditions
$\partial_\nu h^{\mu\nu} = 0$ do not hold unless the evolution equations for
the matter are also satisfied.

The Fokker action is obtained by inserting back
into~\eqref{Sm}--\eqref{Sg} an explicit PN iterated solution of the
field equations~\eqref{EFE} given as a functional of the particle's
trajectories, i.e., an explicit PN metric $g_{\mu\nu}(\bm{x};
\bm{y}_B(t), \bm{v}_B(t), ...)$ at point $\bm{x}$. The extra variables
indicated by ellipsis are higher derivatives such as accelerations
$\bm{a}_B(t)$ or derivatives of accelerations $\bm{b}_B(t)$. Their
presence is due to the fact that we solve Eqs.~\eqref{EFE} without
replacing accelerations because we are off-shell at this stage. Thus,
the Fokker generalized PN action, depending not only on positions and
velocities but also on accelerations and their derivatives, reads
\begin{align}\label{SF}
S_\text{F}\left[\bm{y}_B(t), \bm{v}_B(t), ...\right] &=
\int \ud^{4}x \,\mathcal{L}_g\left[\bm{x};
  \bm{y}_B(t), \bm{v}_B(t), ...\right] \nonumber\\ &-
\sum_A m_A c^{2} \int \ud t \sqrt{-g_{\mu\nu}\left(\bm{y}_A(t);
  \bm{y}_B(t), \bm{v}_B(t), ...\right)
  \,v_A^{\mu}v_A^{\nu}/c^{2}}\,.
\end{align}
Now, by the stationarity of the total action $S=S_g+S_m$ for the PN iterated
solution, the PN equations of motion are nothing but the Euler-Lagrange
equations of $S_\text{F}$ for the particles. Once they have been obtained,
they may then be order reduced as usual, by replacing all accelerations by the
PN equations of motion themselves. The classical Fokker action is completely
equivalent, in the ``tree-level'' approximation, to the effective action used
in the EFT.~\cite{FS4PN,FStail,GLPR16,PR17}

In~\eqref{SF}, the gravitational term integrates over the whole space
a PN solution of the EFE that is valid only in the near zone of the
source. Denoting by $\overline{h}$ the \textit{PN expansion} of the
full-fledged gravitational field $h\equiv h(\bm{x}; \bm{y}_A(t),
\bm{v}_A(t), ...)$, solution of the EFE~\eqref{EFE}, we have the
equality $h=\overline{h}$ in the near zone of the matter system. By
contrast, outside the near zone, $\overline{h}$ is not expected to
agree with $h$ and typically diverges at infinity. On the other hand,
the \textit{multipole expansion} of the metric perturbation, denoted
$\mathcal{M}(h)$, agrees with $h$ in all the exterior region of the
source, but blows up when formally extended inside the near zone as
$r\to 0$.~\cite{BD86} To properly define the Fokker action, we
initially introduced~\cite{BBBFMa} a \textit{Hadamard regularization}
(HR). With that regularization, we demonstrated that the gravitational part
of the Fokker Lagrangian, say $L_g^\text{HR}$, can be written as a
space integral over the looked-for PN Lagrangian density, plus an
extra contribution involving the multipole expansion:
\begin{equation}\label{lemma1}
L_g^\text{HR} = \mathop{\text{{\rm FP}}}_{B=0}\int \ud^{3}\bm{x}
\,\Bigl(\frac{r}{r_0}\Bigr)^B\,\overline{\mathcal{L}}_g +
\mathop{\text{{\rm FP}}}_{B=0}\int \ud^{3}\bm{x}
\,\Bigl(\frac{r}{r_0}\Bigr)^B\mathcal{M}(\mathcal{L}_g)\,.
\end{equation}
Here, we have introduced a regulator $(r/r_0)^B$, with $B$ being a complex
number, and a finite part (FP) at $B=0$ in order to cure the
divergences of the PN expansion when
$r\equiv\vert\bm{x}\vert\to+\infty$ in the first term while dealing
with the singular behaviour of the multipole expansion when $r\to 0$
in the second one. The constant $r_0$, representing an IR scale in
the first term and a UV scale in the second, cancels out between the
two contributions. We have proved, though, that the second term
in~\eqref{lemma1} does not contribute to $S_\text{F}$ below the 5.5PN
order, hence we consider at 4PN order:
\begin{equation}\label{lemma14PN}
L_g^\text{HR} = \mathop{\text{{\rm FP}}}_{B=0}\int \ud^{3}\bm{x}
\,\Bigl(\frac{r}{r_0}\Bigr)^B\,\overline{\mathcal{L}}_g \,.
\end{equation}

\section{The 4PN conservative dynamics}

The general PN solution that matches an exterior solution with
retarded boundary conditions at infinity may be decomposed in two
pieces. The first one consists of the naive near zone expansion of the
retarded integral of the PN source, each term being regularized by
means of the same FP procedure as in Eq.~\eqref{lemma14PN}.
The second piece is a homogeneous multipolar solution regular inside
the source, expanded in the near zone,~\cite{PB02,BFN05}
\begin{equation}\label{hgen}
\overline{h}^{\mu\nu} = \frac{16\pi G}{c^4} \overline{\Box}^{-1}_\text{ret}
\overline{\tau}^{\mu\nu} - 
\frac{2G}{c^4}\sum^{+\infty}_{\ell=0}
\frac{(-)^\ell}{\ell!}\,\partial_L \!\overline{\left\{
\frac{\mathcal{R}^{\mu\nu}_L (t-r/c)-\mathcal{R}^{\mu\nu}_L
  (t+r/c)}{r} \right\}}\,.
\end{equation} 
The multipole moments $\mathcal{R}^{\mu\nu}_L$ in~\eqref{hgen} are
functionals of the multipole expansion of the effective gravitational
source in the EFE, $\mathcal{M}(\tau^{\mu\nu})$, thus depending on the
boundary conditions imposed at infinity.
Most importantly, the functions $\mathcal{R}^{\mu\nu}_L$ are
responsible for the tail effects in the near zone metric. At the 4PN
order, where they first appear in $\overline{h}^{\mu\nu}$, it is
sufficient to consider only quadrupolar tail terms corresponding to
the interaction between the total ADM mass $M$ of the source and its
Symmetric Trace-Free (STF) quadrupole moment $I_{ij}$. Inserting them
into the original Fokker action, we obtain, after redefining the
matter variables $\bm{y}_A$ with the help of an appropriate
non-local-in-time 4PN shift, a net contribution
\begin{equation}\label{StailHad}
S_\text{F}^\text{tail} = \frac{G^2M}{5c^8}
\mathop{\text{Pf}}_{2s_0/c}\int\!\!\!\int \frac{\ud t\ud t'}{\vert
  t-t'\vert} \,I_{ij}^{(3)}(t) \,I_{ij}^{(3)}(t') \,,
\end{equation}
where the upper indices $(3)$ represent third order time differentiation.
With HR, a dependence on the scale $s_0$ [\textit{a priori}
  different from $r_0$ in~\eqref{lemma14PN}] occurs through the
definition of the Hadamard \textit{partie finie} Pf.~\footnote{For any
  regular function $f(t)$ tending to zero sufficiently rapidly when
  $t\to\pm\infty$, we have
$$\mathop{\text{Pf}}_{\tau_0} \int_{-\infty}^{+\infty} \ud
  t'\,\frac{f(t')}{\vert t-t'\vert} \equiv \int_0^{+\infty} \ud\tau
  \ln\left(\frac{\tau}{\tau_0}\right)\left[f^{(1)}(t-\tau) -
    f^{(1)}(t+\tau)\right]\,.$$} Varying the action~\eqref{StailHad}
with respect to the particle's worldlines, we recover the conservative
part of the known 4PN tail effect.~\cite{BD88}

Following previous works on the 3PN equations of
motion~\cite{BFeom,BDE04} we shall proceed in several steps. First, we
parametrize the particular solution
$\overline{\Box}^{-1}_\text{ret}$ in the metric by means of specific
PN potentials. Next, those potentials are computed at any point in
three-dimensional space and inserted into the action. To deal with
quadratic source terms, we extensively make use of the important Fock
function $g=\ln(r_1+r_2+r_{12})$, such that $\Delta g =
r_1^{-1}r_2^{-1}$ with $r_A=\vert\boldsymbol{x}-\boldsymbol{y}_A\vert$
and $r_{12}=\vert\boldsymbol{y}_1-\boldsymbol{y}_2\vert$. We also need
to integrate a cubic source term for which we resort to more
complicated elementary solutions.~\cite{BFP98} The integration of the
Lagrangian density is then implemented by means of a Hadamard
regularization, later corrected to a dimensional regularization (DR)
for the UV divergences. The UV poles $\propto 1/(d-3)$ are then
renormalized through a redefinition of the particle's
worldlines.~\cite{BDE04}

\section{IR divergences and ambiguity parameters at the 4PN order}

The ensuing Fokker Lagrangian~\cite{BBBFMa} depends on the IR length $r_0$ and
on the scale $s_0$ in the tail term~\eqref{StailHad}. However, we have shown
that these two scales combine into a single undetermined constant
$\alpha=\ln (r_0/s_0)$ after suitable shifts of the particle's worldlines.
Then, the tail integral~\eqref{StailHad} simply involves the
separation distance $r_{12}$ as ``partie finie'' scale. The constant $\alpha$
cannot be eliminated and is considered to be an \textit{ambiguity} parameter,
equivalent to the ambiguity parameter called $C$ in the ADM Hamiltonian
formalism.\cite{DJS14,DJS16}

The ambiguity parameters are associated with IR divergences, which are in turn
linked to the presence of the tail effect at the 4PN order.\cite{BD88} It is
thus important to check the ``stability'' of the calculation under a change of
regularization procedure for the IR divergences, and eventually to determine
which regularization should be used. We argued~\cite{BBBFMb} that the Fokker
Lagrangian derived by resorting to dimensional regularization for both IR and
UV divergences is not dynamically equivalent to the HR Lagrangian obtained via
our original approach, the difference being composed of two and only two types
of terms (modulo some irrelevant shifts of the trajectories):
\begin{equation}\label{Lnew}
L^\text{DR} = L^\text{HR} + \frac{G^4 m
  \,m_1^2m_2^2}{c^8r_{12}^4}\Bigl(\delta_1 (n_{12}v_{12})^2 + \delta_2
v_{12}^2 \Bigr)\,,
\end{equation}
where $(n_{12}v_{12})$ denotes the scalar product between the unit
separation vector $\bm{n}_{12}=(\bm{y}_1-\bm{y}_2)/r_{12}$ and the
relative velocity $\bm{v}_{12}=\bm{v}_1-\bm{v}_2$, while $v_{12}^2=
(v_{12}v_{12})$ and $m=m_1+m_2$. A pragmatic way to circumvent the
problem is to acknowledge our (provisional) ignorance about the real
values of $\delta_1$, $\delta_2$ and regard them as ambiguity
parameters. Moreover, the terms containing $\alpha$ in $L^\text{HR}$
can be put precisely in the form of the extra terms in~\eqref{Lnew} so that
$\alpha$ can be absorbed into a redefinition of the two ambiguity
parameters $\delta_1$ and $\delta_2$ without loss of generality.

Now, it turns out that the two ambiguity parameters are
\textit{uniquely fixed} by making our dynamics compatible with
existing gravitational self-force (GSF) calculations of the conserved
energy and periastron advance for circular orbits in the small
mass-ratio limit $\nu=m_1m_2/m^2\to 0$ (see the next section). Nonetheless,
it is important to determine them from \textit{first principles},
i.e., without resorting to external calculations. A recent
progress has been made in that direction:~\cite{BBBFMc} We have
replaced the HR prescription for Eq.~\eqref{lemma14PN} above
by a full DR evaluation based on
\begin{equation}\label{DRprescr}
L_g^\text{DR} = \int \ud^{d}\bm{x} \,\overline{\mathcal{L}}_g \,,
\end{equation}
for the instantaneous terms, and computed the analogue of the tail
term~\eqref{StailHad} in $d=3+\varepsilon$ dimensions. Notably, the
computation of the difference between the two prescriptions for the
instantaneous terms, i.e., $L_g^\text{DR}-L_g^\text{HR}$, is quite lengthy as
it depends on the detailed structure of the expansion of the integrand at
infinity. We proved that, in DR, the instantaneous terms develop an IR pole,
but that it is exactly cancelled by a corresponding UV pole coming from the
tail term in $d$ dimensions (related cancellation of poles has been discussed
in the EFT formalism~\cite{GLPR16,PR17}). Finally, with our full DR
calculation, we found that the two ambiguity parameters can be expressed with
a single parameter $\kappa$ as
\begin{equation}\label{ambparam}
\delta_1 = \frac{1733}{1575} - \frac{176}{15}\kappa \,,\qquad\delta_2
= - \frac{1712}{525} + \frac{64}{5}\kappa \,.
\end{equation}
This parameter $\kappa$ comes from our computation of the tails and is
(provisionally) left undetermined. It is equivalent to our former
parameter $\alpha$ or to the ambiguity parameter $C$ in the Hamiltonian
formalism.~\cite{DJS14} The computation of $\kappa$ from first principles is
in progress.~\cite{MBFB17} Let us now see how to compute it thanks to
the circular orbit limit of the invariants of the motion.

\section{Conserved energy and periastron advance at 4PN order}

To investigate the notions of conserved energy and angular momentum in
the case of a non-local-in-time dynamics, we adopt the Hamiltonian
formalism where the two-body system is described by the canonical
conjugate variables $\bm{y}_A$ and $\bm{p}_A$. The Hamiltonian is made
of a local instantaneous piece (containing many instantaneous terms up
to 4PN order) and the non-local-in-time tail part which is the
analogue of Eq.~\eqref{StailHad}, namely
\begin{equation}\label{Htail}
H^\text{tail}[\bm{x}_A, \bm{p}_A] = -
\frac{G^2M}{5c^8}\,\hat{I}_{ij}^{(3)}(t) \mathop{\text{Pf}}_{2s_0/c}
\int_{-\infty}^{+\infty} \frac{\ud \tau}{\vert\tau\vert}
\,\hat{I}_{ij}^{(3)}(t+\tau) \,.
\end{equation}
The hat over the quadrupole moment means that all time derivatives
must be explicitly evaluated by means of the Newtonian equations of
motion. It is crucial to realize that, in Hamilton's equations, the tail
part of the Hamiltonian is to be differentiated in the sense of
functional derivatives, e.g.,
\begin{equation}\label{dHtail}
\frac{\delta H^\text{tail}}{\delta y^i_A} = -
\frac{2G^2M}{5c^8} \frac{\partial \hat{I}_{jk}^{(3)}}{\partial
    y^i_A}\mathop{\text{Pf}}_{2s_0/c} \int_{-\infty}^{+\infty} \frac{\ud
  \tau}{\vert\tau\vert} \,\hat{I}_{jk}^{(3)}(t+\tau) \,.
\end{equation}
Since the time derivative of $H$ computed on shell is linked to the
partial derivatives of $H$, through the chain rule, the usual
cancellations implying $\ud{H(t)}/\ud t=0$ do not occur when
non-local-in-time contributions are present. We find instead a more
complicated ``non-conservation'' law $\ud{H}/\ud
t=\mathcal{P}^\text{tail}$, where $\mathcal{P}^\text{tail}$ involves
non-local integrals constructed from the tail term~\eqref{Htail}. From
that law, we can derive explicitly the conserved energy $E$
associated with the non-local Hamiltonian.~\cite{BBBFMb} We start by
performing a Taylor expansion of $\hat{I}_{ij}(t+\tau)$ when $\tau \to
0$. To remedy the appearance of divergent integrals, we introduce
in the integrand an exponential cut-off factor
$\ue^{-\epsilon\vert\tau\vert}$ for some $\epsilon>0$, and let
$\epsilon$ tend to zero at the end of our calculation.
It is then straightforward to recast $\mathcal{P}^\text{tail}$ as a
total time derivative, say $-\ud \Delta H^\text{tail}/\ud t$, so that
$E=H+\Delta H^\text{tail}$.
Aiming at getting a non-perturbative (resummed) expression for $\Delta
H^\text{tail}$, we notice that the Newtonian quadrupole moment, being
a periodic function of time, may be conveniently decomposed in
discrete Fourier series, with coefficients ${}_p\mathcal{I}_{ij}$
($p\in \mathbb{N}$). All integrals entering $\Delta H^\text{tail}$ can
be evaluated in closed (albeit Fourier-expanded) form, which yields
\begin{equation}\label{DeltaHtail}
\Delta H^\text{tail} = - \frac{2G^2M \omega^6}{5c^8}
\biggl[\sum_{p}\,\vert\mathop{{\mathcal{I}}}_{p}{}_{\!\!ij}\vert^2 p^6
  - \frac{1}{2} \sum_{p+q \not=
    0}\,\mathop{{\mathcal{I}}}_{p}{}_{\!\!ij}\mathop{{\mathcal{I}}}_{q}{}_{\!\!ij}
  \,\frac{p^3q^3(p-q)}{p+q}
  \ln\left|\frac{p}{q}\right|\,\ue^{\ui(p+q)\ell}\biggr]\,,
\end{equation}
where $\omega$ represents the orbital frequency. Remarkably, this
expression contains a constant (DC) contribution [first term in
Eq.~\eqref{DeltaHtail}] proportional to the gravitational
wave energy flux, $\mathcal{F}^\text{GW}=\frac{G}{5
  c^5}\langle(I^{(3)}_{ij})^2\rangle$. The remaining (AC) terms
average to zero, and are strictly zero in the
case of circular orbits. A similar procedure allows us to construct
the conserved angular momentum.~\cite{BBBFMb}

The complete expression of the energy through 4PN order in the
limiting case of circular orbits is the sum of the instantaneous part of the
4PN dynamics, composed of many different terms, and of the tail part,
composed of~\eqref{Htail} plus the crucial DC contribution
in~\eqref{DeltaHtail}. After reducing to the frame of the center of
mass and specializing to circular orbits, we obtain
\begin{align}\label{Ecirc4PN}
	E &= -\frac{m\nu c^2 x}{2} \biggl\{ 1 + \left( -
        \frac{3}{4} - \frac{\nu}{12} \right) x + \left( - \frac{27}{8}
        + \frac{19}{8} \nu - \frac{\nu^2}{24} \right) x^2 \nonumber
        + \left( - \frac{675}{64} + \biggl[
          \frac{34445}{576} - \frac{205}{96} \pi^2 \biggr] \nu \right. \\ &
     \left. \qquad~ - \frac{155}{96} \nu^2 - \frac{35}{5184} \nu^3 \right) x^3
       + \left( - \frac{3969}{128} +
        \left[-\frac{123671}{5760}+\frac{9037}{1536}\pi^2 +
          \frac{896}{15}\gamma_\text{E}+ \frac{448}{15} \ln(16
          x)\right]\nu\right.\nonumber\\ & \qquad~ \left.+
        \left[-\frac{498449}{3456}+\frac{3157}{576}\pi^2\right]\nu^2
        +\frac{301}{1728}\nu^3 + \frac{77}{31104}\nu^4\right) x^4
        \biggr\} \,.
\end{align}
The PN parameter reads $x=(G m \omega/c^3)^{2/3}$ and we have used the approximation
$M=m_1+m_2$ in the tail terms. We adjust the remaining ambiguity
parameter in~\eqref{ambparam} by comparing to the circular energy obtained in
the GSF framework, at first order in the perturbative expansion in the small
mass ratio limit.~\cite{LBW12,BiniD13} The correct value is
$ \kappa = \frac{41}{60}$.
Such value agrees with the one found in the computation of the tail
term in $d$ dimensions (including both conservative and dissipative
effects) by means of EFT methods.~\cite{GLPR16}

Finally we report the complete expression of the periastron advance at
the 4PN order, for a slightly non-circular orbit, in the limit where the
eccentricity goes to zero:~\cite{DJS15eob,BBBFMb}
\begin{align}\label{Ktotal}
& K = 1 + 3 x + \left(\frac{27}{2} - 7\nu\right) x^2 +
\left(\frac{135}{2}
+\left[-\frac{649}{4}+\frac{123}{32}\pi^2\right]\nu+ 7\nu^2\right) x^3
+ \left(\frac{2835}{8} +\left[-\frac{275941}{360}
    \right. \right. \nonumber \\ & \left. \left. +\frac{48007}{3072}\pi^2
 -\frac{1256}{15}\ln x - \frac{592}{15}\ln 2 - \frac{1458}{5}\ln 3 -
  \frac{2512}{15}\gamma_\text{E}\right]\nu
+ \left[\frac{5861}{12}-\frac{451}{32}\pi^2\right]\nu^2 -
\frac{98}{27}\nu^3\right) x^4\,.
\end{align}
Note that, for the previous value of $\kappa$, the result
agrees directly with GSF calculations. The GSF contribution
to the periastron is generally described by means of the function $\rho(x)$
such that $K^{-2} = 1 - 6 x + \nu\rho(x) + \mathcal{O}(\nu^2)$:
\begin{align}\label{rho}
\rho &= 14 x^2 + \left(\frac{397}{2}-\frac{123}{16}\pi^2\right) x^3
\nonumber\\& + \left(-\frac{215729}{180} + \frac{58265}{1536}\pi^2 +
\frac{1184}{15}\ln 2 + \frac{2916}{5}\ln 3 +
\frac{5024}{15}\gamma_\text{E} + \frac{2512}{15}\ln x\right) x^4\,.
\end{align}
The 4PN coefficient $\rho_\text{4PN}=a_\text{4PN}+b_\text{4PN}\ln x$,
in particular the coefficient $a_\text{4PN}$ with numerical
value $a_\text{4PN}\simeq 64.6406$, is in perfect agreement with GSF
numerical results.~\cite{vdM16} It is worth mentionning that the GSF
periastron advance (analytical or numerical) is not computed directly
but indirectly deduced from the so-called redshift variable via the
first law of binary mechanics, but the latter has been checked to hold
even at the 4PN order for the non-local-in-time dynamics.~\cite{BL17}


\section*{References}


\begin{thebibliography}{30}

\bibitem{BuonSathya15}
 A.~Buonanno and B.S. Sathyaprakash.
\newblock Sources of gravitational waves: Theory and observations.
\newblock In A.~Ashtekar, B.K. Berger, J.~Isenberg, and M.A.H. MacCallum,
 editors, {\em General Relativity and Gravitation: A Centennial Perspective},
 page 513, 2015.

\bibitem{ALBSW15}
S.~Akcay, A.~Le~Tiec, L.~Barack, N.~Sago and N.~Warburton.
\newblock {\em Phys. Rev. D}, 91:124014, 2015.

\bibitem{BuonD99}
 A.~Buonanno and T.~Damour.
\newblock {\em Phys. Rev. D}, 59:084006, 1999.

\bibitem{LD17}
H.A. Lorentz and J.~Droste.
\newblock {\em The motion of a system of bodies under the influence of their
  mutual attraction, according to Einstein's theory}, page 330.
\newblock Nijhoff, The Hague, 1937.
\newblock Versl. K. Akad. Wet. Amsterdam {\bf 26}, 392 and 649 (1917).

\bibitem{EIH}
A.~Einstein, L.~Infeld, and B.~Hoffmann.
\newblock {\em Ann. Math.}, 39:65--100, 1938.

\bibitem{Dhouches} T.~Damour.
  \newblock Gravitational radiation and
  the motion of compact bodies.  \newblock In N.~Deruelle and
  T.~Piran, editors, {\em Gravitational Radiation}, pages 59--144,
  Amsterdam, 1983. North-Holland Company.

\bibitem{Ehletal76}
J.~Ehlers, A.~Rosenblum, J.N.~Goldberg and P.~Havas.
\newblock {\em Astrophys. J.}, 208:L77, 1976.

\bibitem{FS4PN}
S.~Foffa and R.~Sturani.
\newblock {\em Phys. Rev. D}, 87:064011, 2012.

\bibitem{FStail}
S.~Foffa and R.~Sturani.
\newblock {\em Phys. Rev. D}, 87:044056, 2013.

\bibitem{JaraS13}
P.~Jaranowski and G.~Sch{\"a}fer.
\newblock {\em Phys. Rev. D}, 87:081503(R), 2013.

\bibitem{DJS14}
T.~Damour, P.~Jaranowski and G.~Sch{\"a}fer.
\newblock {\em Phys. Rev. D}, 89:064058, 2014.

\bibitem{DJS16}
T.~Damour, P.~Jaranowski, and G.~Sch{\"a}fer.
\newblock {\em Phys. Rev. D}, 93:084014, 2016.

\bibitem{BBBFMa}
L.~Bernard, L.~Blanchet, A.~Boh\'e, G.~Faye and S.~Marsat.
\newblock {\em Phys. Rev. D}, 93:084037, 2016.

\bibitem{BBBFMb}
L.~Bernard, L.~Blanchet, A.~Boh\'e, G.~Faye and S.~Marsat.
\newblock {\em Phys. Rev. D}, 95:044026, 2017.

\bibitem{BBBFMc} L.~Bernard, L.~Blanchet, A.~Boh\'e, G.~Faye and
  S.~Marsat. arXiv:1706:08480, 2017.

\bibitem{GLPR16}
C.~R. Galley, A.~K. Leibovich, R.~A. Porto and A.~Ross.
\newblock {\em Phys. Rev. D}, 93:124010, 2016.

\bibitem{PR17}
R.~A. Porto and I.~Z. Rothstein. arXiv:1703.06433, 2016.

\bibitem{BD86}
L.~Blanchet and T.~Damour.
\newblock {\em Phil. Trans. Roy. Soc. Lond. A}, 320:379--430, 1986.

\bibitem{PB02}
O.~Poujade and L.~Blanchet.
\newblock {\em Phys. Rev. D}, 65:124020, 2002.

\bibitem{BFN05}
L.~Blanchet, G.~Faye and S.~Nissanke.
\newblock {\em Phys. Rev. D}, 72:044024, 2005.

\bibitem{BD88}
L.~Blanchet and T.~Damour.
\newblock {\em Phys. Rev. D}, 37:1410, 1988.

\bibitem{BFeom}
L.~Blanchet and G.~Faye.
\newblock {\em Phys. Rev. D}, 63:062005, 2001.

\bibitem{BDE04}
L.~Blanchet, T.~Damour, and G.~Esposito-Far{\`e}se.
\newblock {\em Phys. Rev. D}, 69:124007, 2004.

\bibitem{BFP98}
L.~Blanchet, G.~Faye, and B.~Ponsot.
\newblock {\em Phys. Rev. D}, 58:124002, 1998.

\bibitem{MBFB17}
T.~Marchand, L.~Blanchet, G.~Faye, and L.~Bernard.
\newblock In preparation, 2017.

\bibitem{LBW12}
A.~Le~Tiec, L.~Blanchet and B.~Whiting.
\newblock {\em Phys. Rev. D}, 85:064039, 2012.

\bibitem{BiniD13}
D.~Bini and T.~Damour.
\newblock {\em Phys. Rev. D}, 87:121501(R), 2013.

\bibitem{DJS15eob}
T.~Damour, P.~Jaranowski, and G.~Sch{\"a}fer.
\newblock {\em Phys. Rev. D}, 91:084024, 2015.

\bibitem{vdM16}
M.~van de Meent.
\newblock {\em Phys. Rev. Lett.}, 118:011101, 2017.

\bibitem{BL17}
A.~Le~Tiec and L.~Blanchet. arXiv:1702.0639, 2017.

\end{thebibliography}
\end{document}